\documentclass[twocolumn,twocolappendix]{aastex631} %

\usepackage{multirow}
\usepackage{gensymb}
\usepackage{float}
\usepackage{subfigure}
\usepackage{graphicx}
\usepackage{threeparttable}
\usepackage{amsmath}
\usepackage{color}
\usepackage{mathrsfs}
\usepackage{textcomp, gensymb}
\newcommand{\HI}{H\,\textsc{i}}

\newcommand{\Ha}{\mbox{H$\alpha$}}

\newcommand{\Ms}{{\mbox{$M_*$}}}
\newcommand{\Md}{{\mbox{$M_\mathrm{dust}$}}}
\newcommand{\Mg}{{\mbox{$M_\mathrm{gas}$}}}

\newcommand{\um}{{\mbox{$\mu$m}}}
\newcommand{\usurf}{{\mbox{$\mathrm{mag\,arcsec^{-2}}$}}}

\newcommand{\sigmagas}{\mbox{$\Sigma_{\rm gas}$}}
\newcommand{\sigmasfr}{\mbox{$\Sigma_{\rm SFR}$}}
\newcommand{\sigmadust}{\mbox{$\Sigma_{\rm dust}$}}

\shorttitle{Star Formation in Nearby Galaxy Mergers}
\shortauthors{Li, Ho \& Shangguan}

\begin{document}
\title{The Subtle Effects of Mergers on Star Formation in Nearby Galaxies}

\correspondingauthor{Yang A. Li; Jinyi Shangguan}
\email{liyang23@pku.edu.cn; shangguan@mpe.mpg.de}

\author[0000-0002-3309-8433]{Yang A. Li}
\affiliation{Kavli Institute for Astronomy and Astrophysics, Peking University, Beijing 100871, P. R. China}
\affiliation{Department of Astronomy, School of Physics, Peking University, Beijing 100871, P. R. China}

\author[0000-0001-6947-5846]{Luis C. Ho}
\affiliation{Kavli Institute for Astronomy and Astrophysics, Peking University, Beijing 100871, P. R. China}
\affiliation{Department of Astronomy, School of Physics, Peking University, Beijing 100871, P. R. China}

\author[0000-0002-4569-9009]{Jinyi Shangguan}
\affil{Max-Planck-Institut f\"{u}r extraterrestrische Physik, Gie{\ss}enbachstr. 1, D-85748 Garching, Germany}

\begin{abstract}
Interactions and mergers play an important role in regulating the physical properties of galaxies, such as their morphology, gas content, and star formation rate (SFR). Controversy exists as to the degree to which these events, even gas-rich major mergers, enhance star formation activity. We study merger pairs selected from a sample of massive ($M_* \ge 10^{10}\,M_\odot$), low-redshift ($z = 0.01-0.11$) galaxies located in the Stripe~82 region of the Sloan Digital Sky Survey, using stellar masses, SFRs, and total dust masses derived from a new set of uniformly measured panchromatic photometry and spectral energy distribution analysis. The dust masses, when converted to equivalent total atomic and molecular hydrogen, probe gas masses as low as $\sim 10^{8.5}\,M_\odot$. Our measurements delineate a bimodal distribution on the $M_{\rm gas}-M_*$ plane: the gas-rich, star-forming galaxies that trace the well-studied gas mass main sequence, and passive galaxies that occupy a distinct, gas-poor regime. These two populations, in turn, map into a bimodal distribution on the relation between SFR and gas mass surface density. Among low-redshift galaxies, galaxy mergers, including those that involve gas-rich and nearly equal-mass galaxies, exert a minimal impact on their SFR, specific SFR, or star formation efficiency. Starbursts are rare. The star formation efficiency of gas-rich, minor mergers even appears suppressed. This study stresses the multiple, complex factors that influence the evolution of the gas and its ability to form stars in mergers.
\end{abstract}


\section{Introduction}

Interactions and mergers play an essential role in galaxy evolution in a hierarchical Universe \citep{White1991}. Apart from secular evolution through accreting gas from filaments \citep{Dekel2009a}, galaxies can increase significantly their stellar mass ($M_*$) and halo mass through merging with other galaxies \citep{Blumenthal1984CDM, Faber2007, Neistein2008, Cattaneo2011, RodriguezGomez2016}, as well as drastically alter their structure through morphological transformation \citep{Toomre1972, Mihos1996, Lotz2010, Hambleton2011}. Observations support the picture that merger-induced instabilities drive gas flows into the center of merger remnants \citep{Blumenthal2018}, dilute the gas-phase metallicity \citep{Rupke2010, Torrey2012, Bustamante2018}, and enhance the star formation rate (SFR) on either nuclear or global scales \citep{Ellison2008, Ellison2010, Robotham2014, Xu2021}. Numerical simulations of galaxy pairs, which provide an idealized framework to elucidate the merger process and its influence on the galaxy members, confirm that galaxy interactions and mergers enhance the SFR \citep{DiMatteo2007, Cox2008, Scudder2012, Perret2014, Moreno2015, Moreno2021} and regulate the gas content \citep{Mihos1996, Sinha2009, Sparre2022} in the nuclear region \citep{Moreno2021}, as well as set the stage to facilitate the ignition of black hole activity \citep{DiMatteo2005, Hopkins2006, Capelo2015, Capelo2017, ByrneMamahit2022}. The growth of the central black hole and bulge, in turn, can quench star formation \citep{Springel2005Nature, Hopkins2009ApJ}.

Mergers are often invoked as the triggering mechanism of starbursts. Tidal compression and shocks induced by mergers enhance gas density \citep{Barnes2004, Schweizer2009, Renaud2014} to the point that the star formation efficiency can be high enough to qualify the galaxy as a starburst, such as those witnessed in luminous and ultraluminous infrared galaxies \citep{Sanders1996, Shangguan2019}. However, the nature of star formation activity in merging galaxies is far from clear-cut. A wide range of star formation activity exists. Some studies find strong SFR enhancement in closely projected merger pairs \citep{Scudder2012, Davies2015GAMA, Garduno2021} or in pairs of large separation \citep{Ellison2010, Patton2016}, while others conclude that SFR is enhanced only moderately \citep{Xu2010, Knapen2009, Knapen2015, BarreraBallesteros2015}, typically by a factor of $\sim 2$ \citep{Lin2007, Ellison2013, Pan2018, Silva2018} or less \citep{Wong2011, Pearson2019}. \citet{Knapen2015} even suggest that a considerable fraction of merging galaxies, quite contrary to naive expectations, exhibit a reduced level of star formation compared to non-mergers. Mergers may fail to lead to starbursts because the timescale for the close encounter or coalescence needed to trigger star formation enhancement \citep[$\sim 10-100\,\rm Myr$; ][]{DiMatteo2008, Cortijo-Ferrero2017, Pearson2019} is very short compared to the period between encounters \citep{Moreno2019}. Even galaxy pairs with small separation ($<15\,\rm kpc$) may be in a stage that is still too early to boost star formation \citep{Silva2018}. Simulations by \citet{Segovia2022} and \citet{Renaud2022} suggest that starbursts only occur after the disk has had time to form, when the rotational motion of the gas overtakes its velocity dispersion. Although mergers induce strong perturbations, the level of star formation in the galaxy may saturate due to feedback processes in a clumpy, turbulent medium \citep{Perret2014}.

Many previous studies have attempted to identify the main factors that govern the star formation variance in mergers, including the stellar mass ratio of the two galaxies, their initial orbital parameters, projected separation, total stellar mass, bulge-to-total mass ratio, and environment \citep{DiMatteo2007, Cox2008, Ellison2008, Scudder2012, Torrey2012, Perret2014, Domingue2016, Pan2018, Qiu2020, Xu2021}. The stellar mass ratio of the merger pair ($f_{M_*} = M_{*,1}/M_{*,2}$, with $M_{*,1} < M_{*,2}$) is regarded as an essential parameter, one used to separate major from minor mergers, traditionally by $f_{M_*} = 0.25$ \citep{Robotham2013, Davis2015, Pan2018, QYu2022}. Designating the more massive member as the primary galaxy and the less massive one as the secondary, \citet{Davies2015GAMA} find that the SFR is enhanced in all primary galaxies, both in major and minor mergers. By comparison, star formation is suppressed in the secondary component of minor mergers, presumably because the gas is heated or stripped in lower-mass galaxies. However, even in this instance the outcome can be complicated. On the one hand, \citet{DiMatteo2007} show that in merger pairs the less massive galaxy experiences a stronger tidal effect. On the other hand, the high-resolution simulations of \cite{Du2019} demonstrate that while a large fraction of the gas from the less massive galaxy is stripped by ram pressure, the residual gas, under the combined effects of ram pressure confinement and tidal compression at pericentric passages, can sustain bursty central star formation. This may offer an alternative explanation to the observations by \citet{Silva2018}, that the less massive member in a merger pair experiences greater star formation enhancement, which the authors attribute to gas content: splitting their sample into dust-rich (wet), mixed, and dust-poor (dry) mergers according to the colors of the individual galaxy components, they find a higher fraction of starbursts among the dusty, low-mass, star-forming galaxies than in the high-mass members.

Indeed, gas content is considered another critical factor that influences the outcome of star formation in mergers, at least from a theoretical point of view (e.g., \citealt{Bournaud2011, Blumenthal2018}; but see \citealt{Perret2014}). The observational situation, however, again is rather muddled. While many studies point to a direct causal connection between gas content and star formation enhancement in mergers (e.g., \citealt{Casasola2004, Ellison2018, Moreno2019, Moreno2021}), others disagree (e.g., \citealt{Hibbard1996, Ellison2015, Zuo2018, QYu2022}). For example, while \citet{Pan2018} find that in galaxy pairs of comparable stellar mass, large enhancements of SFR are associated only with galaxies that have a significant molecular gas fraction, \cite{Scudder2015} report an inverse correlation between SFR enhancement and gas fraction, an effect that they attribute to the dominance of the strong underlying dependence of SFR on initial gas fraction. These ambiguities can perhaps be ascribed to the difficulty of measuring the gas content of mergers, considering the limited samples that have gas observations \citep{Ellison2015, Ellison2018, Zuo2018, Xu2021, QYu2022} or the potential confusion due to blending by a large telescope beam \citep{Casasola2004}. 

Lastly, we note an often less appreciated complication. It is nontrivial to design a proper control sample to which the mergers need to be compared in order to draw statistically meaningful conclusions about whether and how star formation has been affected by the merging process. A common practice is to create a control sample for each interacting galaxy and then compare its physical properties with the median properties of the control sample. However, most mergers that have gas measurements are star-forming galaxies, whereas their counterparts in the control sample include both star-forming and passive galaxies, if the matching is based only on stellar mass, redshift, and/or environment \citep{Davies2015GAMA, Ellison2018, Qiu2020},  and additionally morphology \citep{Knapen2015} or bulge-to-total mass ratio \citep{He2022BTr}. Alternatively, one can construct a pool of star-forming galaxies to define the control sample \citep{Ellison2013, Scudder2015, Quai2021}, but the simultaneous matching of multiple physical parameters reduces the control sample to a very small number ($\leq 5$) galaxies. The mutual dependence of many galaxy parameters further frustrates any attempt to define a meaningful ``control'' sample.

In this paper, we investigate the effect of tidal interactions on the star formation activity and gas content of galaxies, by comparing the SFR, \Mg, and specific SFR (${\rm sSFR} \equiv {\rm SFR}/M_*$) and star formation efficiency (${\rm SFE} \equiv {\rm SFR}/M_{\rm gas}$) of mergers with non-mergers, where both populations are drawn from a large sample of nearby galaxies selected from the Stripe~82 region \citep{Annis2014} of the Sloan Digital Sky Survey (SDSS; \citealt{York2000}), whose panchromatic, far-ultraviolet (FUV) to far-infrared (FIR) spectral energy distribution (SED) and physical parameters have been analyzed uniformly in the companion work of \cite{Li2023}. Section~\ref{sec_sample} introduces the parent sample, the selection criteria for the merger subsample, as well as the derivation of the physical parameters studied in this work. We present the results in Section~\ref{sec_results}, discuss their implications in Section~\ref{sec_discussion}, and briefly summarize our main conclusions in Section~\ref{sec_summary}. We adopt a $\Lambda$CDM cosmology with $H_0=70\,\rm km\,s^{-1}\,Mpc^{-1}$ and $\Omega_{\Lambda}=0.7$. The stellar masses and SFRs are based on the stellar initial mass function of \citet{Chabrier2003}.

\section{Data}\label{sec_sample}

\subsection{Sample Definition}

We make use of the local galaxies in the SDSS Stripe~82 region that are covered by the Herschel Stripe~82 Survey (HerS; \citealt{VieroHerS2014}) with the SPIRE instrument \citep{Griffin2010} onboard Herschel \citep{Pilbratt2010}. The parent sample comes from the catalog produced by \cite{Li2023}, which is defined by the redshift range of $z = 0.01-0.11$ and a stellar mass cut of $M_* \ge 10^{10}\,M_{\odot}$, using values from the GALEX-SDSS-WISE Legacy Catalog~2 (GSWLC-2; \citealt{Salim2018}). The lower limit on redshift excludes very nearby galaxies, which can have unreliable distances based on redshift, the overall narrow redshift range ensures negligible cosmic evolution in global galaxy properties, and the stellar mass cut yields a sample with high completeness. The parent sample of 2781 galaxies consists of 2668 galaxies with no broad emission lines and 17 broad-line active galactic nuclei, as well as 96 galaxies that do not satisfy the stellar mass cut but are retained because they are companions to a primary galaxy that does.

Interacting galaxies are defined as close pairs with a projected separation of $\Delta\,r<50$ kpc, a physical scale that best captures galaxy pairs during periods of the encounter and before the final coalescence \citep{Ellison2010, CalderonCastillo2019}. Galaxies in those periods have distinct properties compared to isolated galaxies \citep{Moreno2019, Moreno2021}. We impose a radial velocity difference of $\Delta v<600\,\mathrm{km\,s^{-1}}$ to select gravitationally bounded pairs. Some galaxies in the parent sample have a projected close companion that, despite lacking a spectroscopic redshift, exhibits an obvious tidal feature between them that strongly suggests an interaction. We require that the photometric redshift of the close companion, within its uncertainty, be consistent with the spectroscopic redshift of the primary galaxy. Galaxies that lack a spectroscopic redshift are assumed to have a redshift identical to that of their merger companion. 

Among the 2685 galaxies in our parent sample, more than 300 have at least one physically associated companion. We omit systems with three or more objects or multi-component systems that involve the brightest galaxy in a cluster center, as our primary focus is to investigate the influence on star formation induced by companions with different physical properties. Systems that involve the participation of more than one companion or that are situated in the complex environment of cluster centers make it ambiguous to disentangle the influence of individual members. While the primary galaxy must be a member of our parent sample and have $M_* \ge 10^{10}\,M_{\odot}$, the secondary galaxy is allowed to be less massive than this stellar mass limit in order to better explore the effect of minor mergers, although we do not consider galaxy pairs with stellar mass ratios $f_{M_*} < 0.01$ to minimize confusion with bright galactic substructures. Stellar mass ratios are evaluated by taking into account the $3\sigma$ uncertainty of $M_*$, which is $\sim 0.15$~dex \citep{Li2023}. The following analysis involves 174 merging galaxies and a control sample of 2244 isolated galaxies.

\subsection{Photometry and Physical Parameters}\label{sec_property}

\cite{Li2023} presented detailed panchromatic photometry of the sample galaxies, covering 17 bands from the FUV (1539~\AA) to the FIR (500\,\um). We developed a new method to measure matched-aperture photometry consistently for all the bands, in order to obtain robust total flux measurements of relatively bright, nearby galaxies that are partly or fully resolved. Special care was devoted to removing contamination from foreground stars, and simultaneous two-dimensional image decomposition was performed using {\tt GALFITM} \citep{Haubler2013GALFITM} to deblend properly interacting/merging galaxies.  Two-dimensional decomposition is performed simultaneously with images from the FUV to $4.6\,\mu$m, by assuming that the structural parameters change continuously across the wavelengths. This procedure is physically well motivated because these bands are dominated by the stellar emission of the galaxy. The bands at longer wavelengths (W3, W4, and Herschel bands), which are dominated by dust emission, are more challenging to decompose with {\tt GALFITM} because of their limited spatial resolution. We collect the photometric measurements of Herschel observations from the Herschel Extragalactic Legacy Project (HELP; \citealt{Shirley2021HELP}).  Fortunately, HELP provides the decomposed emission of most of our galaxy pairs using a Bayesian-based deblending method that relies on prior information from higher-resolution images in either the optical or near-infrared. HELP provides the global flux of a small fraction of our galaxy pairs whose decomposition failed. For these targets in the Herschel bands and all the galaxy pairs in W3 and W4, we consider the global flux as the flux upper limit of each component in their SED fitting.  Extensive mock experiments helped to establish reliable uncertainties, both statistical and systematic, and to estimate upper limits. We employed the code {\tt CIGALE} \citep{Boquien2019CIGALE} to fit the broadband SED to estimate the global SFR, \Ms, and dust mass (\Md), the latter based on the dust emission model of \cite{DL07}. {\tt CIGALE} gives the Bayesian probability distribution for the model parameters. We consider flux upper limits in calculating the likelihoods of the SED models, and we use mock tests to correct the physical parameters for systematic biases that may arise from photometry of low signal-to-noise ratio (see details in Section 4.4 of \citealt{Li2023}). We exclude a minority of galaxies that have reduced $\chi^2>10$ in the SED fits, which typically stem from sources that are decomposed poorly by {\tt GALFITM} in some near-infrared bands. 

Our ensuing analysis relies critically on having access to a uniform set of reliable gas masses for the merger galaxies and the control sample of non-merger galaxies. Whereas most galaxy studies measure neutral atomic hydrogen through the \HI\ 21~cm line and molecular hydrogen through the rotational transitions of CO, here we adopt the more expedient, indirect method of probing the cold interstellar medium from the total dust mass derived from the infrared continuum emission. This alternative method offers several advantages. While extensive blind extragalactic \HI\ 21~cm surveys have been conducted (e.g., \citealt{Koribalski2004, Haynes2018}) and catalogs of source parameters thereof have been published (e.g., \citealt{Springob2005, Yu2022}), the current samples are limited thus far to relatively nearby ($z \lesssim 0.1$), gas-rich galaxies. Large, unbiased samples of galaxies with CO observations are rarer still (\citealt{Boselli2014, Saintonge2017}; see review in \citealt{Saintonge2022}). On the other hand, relatively deep, homogeneous, and unbiased blind surveys of the FIR continuum from dust emission have been conducted using Herschel, among them HerS \citep{VieroHerS2014}, offering an opportunity to secure the total dust, and hence gas, content of large samples of galaxies. Moreover, even when CO observations are available, the conversion factor from CO to $\rm H_2$ remains a perennial topic of debate \citep{Bolatto2013}. The dust-based method of deriving total gas mass sidesteps this issue. 

We convert the total dust mass to total gas mass using the empirical relation between gas-phase metallicity and gas-to-dust mass ratio ($\delta_{\mathrm{GDR}}$; \citealt{Leroy2011}). From the calibration of \cite{Magdis2012},

\begin{align}\label{equ_GDR}
\mathrm{log}\,\delta_{\mathrm{GDR}} = & (10.54\pm 1.0) - (0.99\pm 0.12) \nonumber \\ 
& \times [12+\mathrm{log(O/H)}],
\end{align}

\noindent
where $12+\mathrm{log(O/H)}$ is the oxygen abundance empirically defined by \citet{PettiniPagel2004}, which can be estimated from the stellar mass-metallicity relation of local galaxies \citep{KewleyEllison2008},

\begin{align}\label{equ_metal}
12+\mathrm{log(O/H)} & =  23.9049 - 5.62784 \, \mathrm{log\,}M_*  \nonumber \\
& + 0.64514\, (\mathrm{log\,}M_*)^2 - 0.02351\,  (\mathrm{log\,}M_*)^3.
\end{align}

\noindent
Following \cite{Shangguan2018}, we add 0.23 dex to $\delta_{\mathrm{GDR}}$ to account for the empirical difference between dust-converted gas masses and direct gas observations. The total uncertainty of the gas masses,  $\sim 0.45$~dex, stems from the scatter of 0.15~dex in $\delta_{\rm GDR}$ propagated from the scatter of the stellar mass-metallicity relation and the typical uncertainty of 0.42~dex for $M_d$ \citep{Li2023}, which principally relates to uncertainties of the dust temperature and absorption coefficient. Dust radiation is very sensitive to temperature; even a tiny change in temperature can induce a large variation in dust mass \citep{Lv2018}.

\begin{figure*}
\centering
\includegraphics[width=0.9\textwidth]{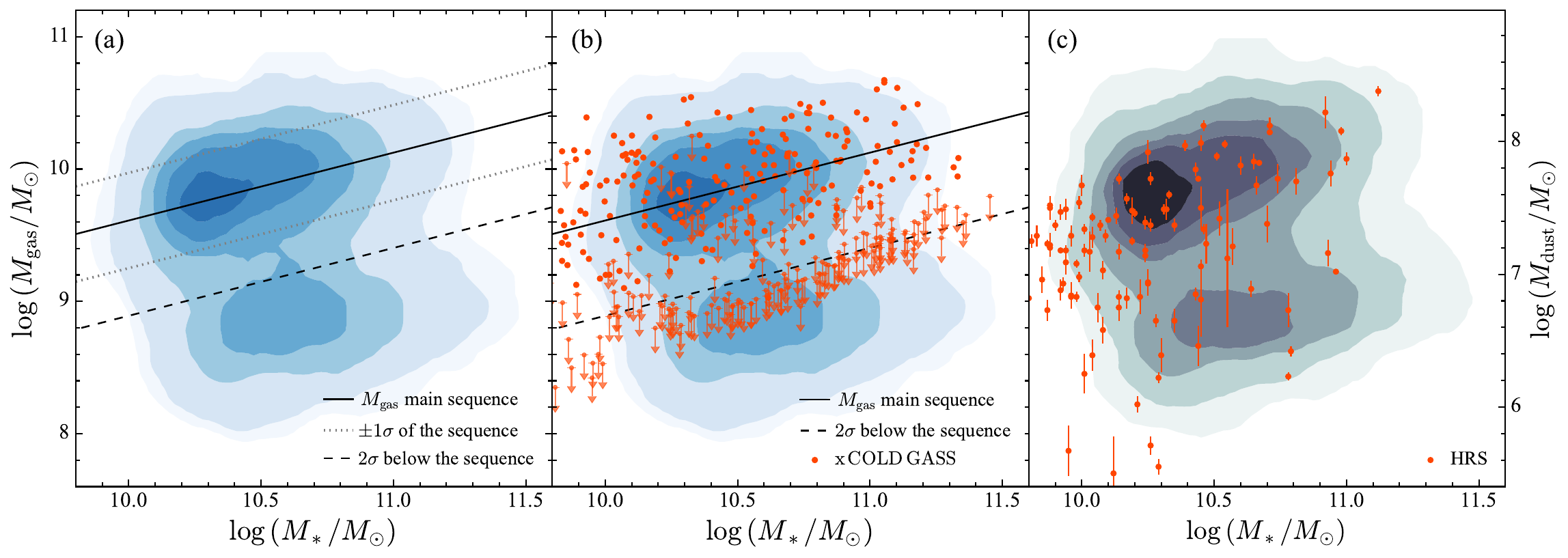}
\caption{Two-dimensional distribution of the total gas and dust mass versus stellar mass. In panels (a) and (b), the background contours delineate the $M_\mathrm{gas} - M_*$ distribution of our sample; in panel (b) the colored points are taken from xCOLD GASS \citep{Saintonge2017}. The black solid line indicates the best fit for the gas mass main sequence, which has a $1\sigma$ scatter of 0.38~dex (gray dotted lines). The threshold for gas-rich objects is given by the black dashed line, which is set to $2\sigma$ below the gas mass main sequence. Panel (c) shows the $M_\mathrm{dust} - M_*$ distribution for our sample in contours, while the red points represent HRS galaxies with dust masses derived by \cite{Ciesla2014}.
}
\label{fig_Mgas_GMMS}
\end{figure*}

\section{Results}\label{sec_results}
\subsection{Total Gas Mass as a Function of Stellar Mass}

We derive total gas masses for the local galaxies in the Stripe~82 field based on their cold dust masses estimated from SED fitting (Section~\ref{sec_property}). Our panchromatic SED fitting includes upper limits in all bands, which allows us to measure meaningful, unbiased dust masses down to very low levels, significant uncertainties notwithstanding. This approach makes it possible to probe the gas mass distribution of gas-poor galaxies, down to $M_{\rm gas} \approx 10^{8.5}\,M_\odot$ (Figure~\ref{fig_Mgas_GMMS}a). The star-forming and passive galaxies form two distinct peaks on the $M_{\rm gas}-M_*$ diagram, separated by a clear gap between them. This distribution qualitatively resembles the bimodal distribution of the star-forming main sequence on the ${\rm SFR}-M_*$ diagram (e.g., \citealt{Noeske2007, Peng2010}). The upper part of the $M_{\rm gas}-M_*$ diagram has been studied in the literature mainly through molecular gas measurements and is often called the gas mass main sequence (e.g., \citealt{Saintonge2016, Bolatto2017, Lin2019, Baker2023}).

\begin{figure*}
\centering
\includegraphics[width=0.85\textwidth]{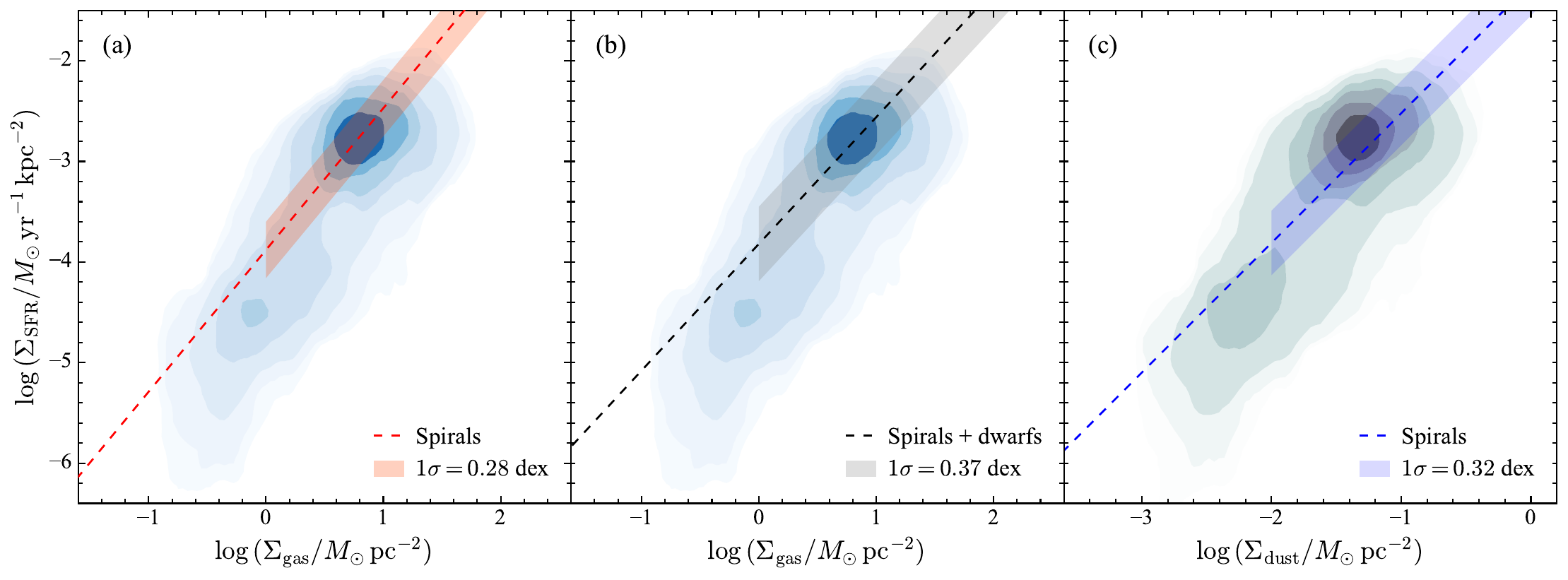}
\caption{Kennicutt-Schmidt relation on the (a and b) \sigmasfr--\sigmagas\ and (c) \sigmasfr--$\Sigma_{\rm dust}$ diagram, where $\Sigma_{\rm dust}$ is the dust mass surface density. The distribution for our sample is plotted as the background contours. The dashed lines in the panels are (a) the K-S law from \cite{delosReyes2019} for spiral galaxies: $\mathrm{log\,}\sigmasfr = 1.41\,\mathrm{log\,}\sigmagas - 3.88$, with a scatter of 0.28~dex; (b) the K-S law from \cite{delosReyes2019} for the combination of spiral and dwarf galaxies: $\mathrm{log\,}\sigmasfr = 1.26\,\mathrm{log\,}\sigmagas - 3.82$, with a scatter of 0.37~dex; and (c) the relation from \cite{Kennicutt2021}: $\mathrm{log\,}\sigmasfr = 1.29\,\mathrm{log\,}\sigmadust - 1.23$, with a scatter of 0.32 dex. All the values of \sigmasfr\ are converted to the \cite{Chabrier2003} initial mass function.}
\label{fig_KSlaw_compare}
\end{figure*}

\begin{figure}[b]
\centering
\includegraphics[width=0.42\textwidth]{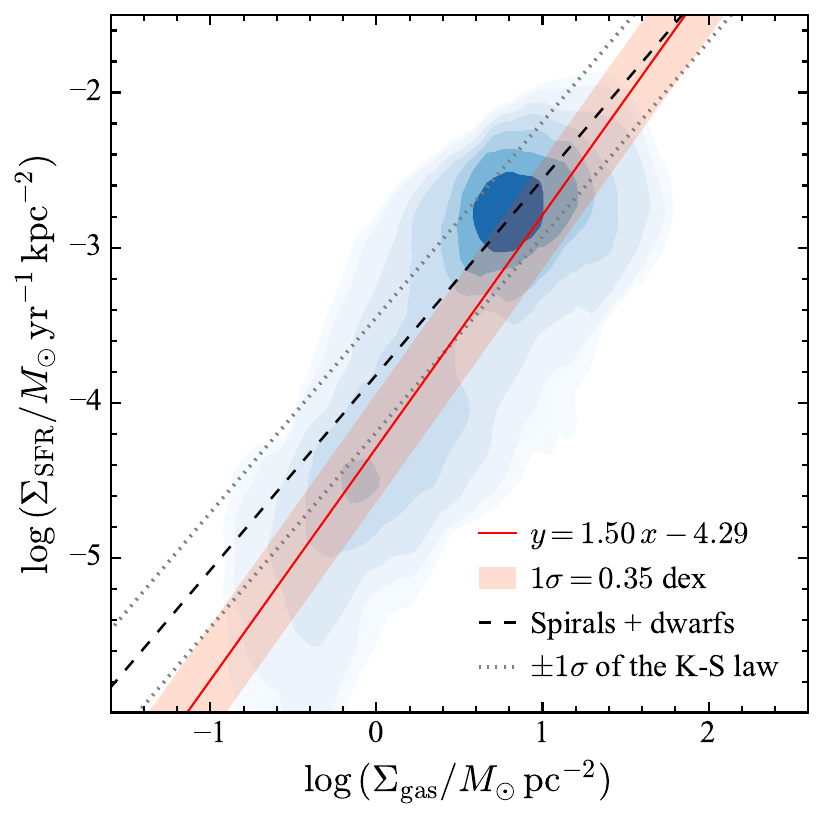}
\caption{Distribution of our sample on the \sigmasfr--\sigmagas\ diagram. The red line shows the linear fit for the entire sample, weighted by the uncertainties. The black dashed line is the fit by \cite{delosReyes2019} for spiral and dwarf galaxies, which has a $1\sigma$ scatter of 0.37~dex (gray dotted lines).}
\label{fig_KSlaw_myfit}
\end{figure}

Following the criterion of \cite{Saintonge2016} and \cite{Li2023} to select gas-rich, star-forming galaxies, we choose galaxies with gas fraction $f_\mathrm{gas} \equiv M_\mathrm{gas}/M_* > 0.05$ to define the gas mass main sequence. Averaging the gas mass in stellar mass bins of width $\Delta \log M_* = 0.15$~dex, we use the {\tt Python} function {\tt curve\_fit} \footnote{\url{https://docs.scipy.org/doc/scipy/reference/generated/scipy.optimize.curve\_fit.html}} from the package {\tt scipy} \citep{Virtanen2020scipy} to fit the binned data with the linear relation

\begin{align}
\mathrm{log}\,M_{\mathrm{gas}} = 0.5127 \, \log\,(\Ms/10^{10}\,M_\odot) + 9.611,
\end{align}

\noindent
which has a $1\sigma$ total scatter of 0.36~dex. The limited dynamic range of stellar mass in our sample does not justify a higher-order fit. We do not attempt to correct for sample completeness because the main goal of this study is simply to construct a sample distribution that can serve as the control sample with which to compare the merger sample. We classify the gas content of the galaxies in our sample based on their distance to the gas mass main sequence. We define gas-rich galaxies as those that lie above the dashed line in Figure~\ref{fig_Mgas_GMMS}a, which is $2\sigma$ below the best-fit main sequence relation of Equation~3. Gas-poor galaxies sit below the dashed line.

To our knowledge, the lower peak of the $M_{\rm gas}-M_*$ distribution has never been reported before. Direct observations of gas content in galaxies through blind surveys of the \HI\ 21~cm or CO lines generally lack the sensitivity to detect large numbers of gas-poor galaxies. For example, the total cold gas (neutral atomic and molecular hydrogen) masses of the xCOLD~GASS survey \citep{Saintonge2017}, to date the largest unbiased sample of $\sim 500$ nearby galaxies with dedicated, relatively deep \HI\ and CO observations, largely populate the upper peak of the $M_{\rm gas}-M_*$ diagram; most of the gas-poor galaxies are upper limits and fail to delineate the lower peak (Figure~\ref{fig_Mgas_GMMS}b). Dust masses are available for larger samples of galaxies, but such measurements are biased because they derive from studies that usually collect only detections from different databases to construct the SED (e.g., \citealt{Dunne2011, Berta2016}). Dust masses are not given for galaxies without FIR detections. In contrast, \cite{Li2023} showed that the dust mass can be reasonably estimated when FIR upper limits are included in the broadband SED fit, even if the uncertainties are considerable. This point is demonstrated in detail using mock SEDs in Section~4.4 of \cite{Li2023}. Targetted Herschel observations of local galaxies, such as the Herschel Reference Survey (HRS) for about 300 galaxies at 15--25~Mpc \citep{Boselli2010}, can provide deep enough detection of cold dust in passive galaxies \citep{Ciesla2014}. As shown in Figure~\ref{fig_Mgas_GMMS}c, the dust mass measurements of the HRS sources are well consistent with the sample distribution in this work. The passive galaxies in HRS, although limited in number, overlap with the lower locus of our passive galaxies, which supports the robustness of the dust masses of our passive galaxies. As discussed in Section~\ref{sec_result_kslaw}, the new dust (gas) masses of the gas-poor galaxy population obey a relation between the surface density of SFR and gas mass that closely follows the extrapolation of the well-studied Kennicutt-Schmidt (K-S: \citealt{Schmidt1959, Kennicutt1998}) law for gas-rich, star-forming galaxies. This, too, lends confidence that the dust (gas) masses of the gas-poor subsample do yield physically meaningful constraints on the cold interstellar medium of this seldom-probed region of parameter space.

\begin{figure*}[t]
\centering
\includegraphics[width=0.98\textwidth]{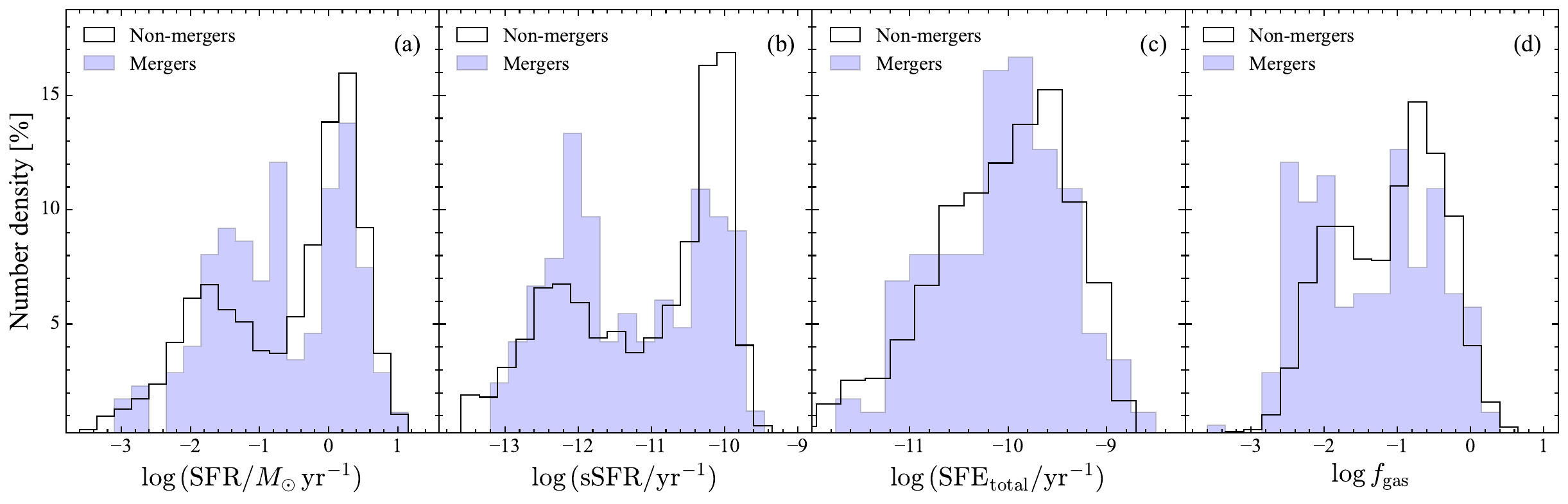}
\caption{Distribution of (a) SFR, (b) sSFR, (c) SFE, and (d) gas-to-stellar mass ratio $f_{\rm gas}$. Mergers and non-mergers are denoted as filled and open histograms, respectively.}
\label{fig:histAll}
\end{figure*}

\subsection{The Kennicutt-Schmidt Relation for Star-forming and Passive Galaxies}
\label{sec_result_kslaw}

The availability of relatively sensitive gas mass estimates and SFRs for a large, homogeneously studied sample of nearby galaxies affords us the opportunity to revisit the star formation relation as commonly expressed using the K-S law. Previous works, mainly based on direct gas measurements, found power-law relations between the surface density of SFR and the surface density of the total or molecular gas mass \citep{Kennicutt1997,Leroy2008,Martig2013,delosReyes2019},

\begin{align}\label{equ_KSmodel}
\sigmasfr = \alpha \, \sigmagas ^{\beta},
\end{align}

\noindent
where $\beta$ is the power-law slope that typically varies from 0.8 to 2, depending on the sample and measurements, and $\alpha$ is the normalization. As in \cite{Kennicutt2021}, we calculate the surface densities \sigmasfr\ and \sigmagas\ by dividing SFR and \Mg, respectively, by $\pi a^2$, where $a$ is the semimajor axis of the galaxy stellar disk, which we approximate using $a_{25}$, the isophotal radius at 25~\usurf\ in the $r$ band, as listed in the SDSS Stripe~82 database.

As with the distribution of $M_{\rm gas}$ versus $M_*$ (Figure~\ref{fig_Mgas_GMMS}), the $\sigmasfr - \sigmagas$ diagram also exhibits a bimodal distribution (Figure~\ref{fig_KSlaw_compare}). The first peak, centered roughly at $\log\, (\sigmagas/M_\odot \, \rm pc^{-2}) \approx 0.8$ and $\log\, (\sigmasfr/M_\odot \, \rm yr^{-1}\,kpc^{-2}) \approx -2.7$, pertains to the group of gas-rich, star-forming galaxies commonly studied in the context of the K-S relation. We draw attention to the gas-poor, passive galaxies clustered around $\log\, (\sigmagas/M_\odot \, \rm pc^{-2}) \approx -0.4$ and $\log\, (\sigmasfr/M_\odot \, \rm yr^{-1}\,kpc^{-2}) \approx -4.7$. The entire distribution of the Stripe~82 galaxies is approximately consistent with a single relation that closely resembles the K-S law, although the gas-poor population may present a mild break below $\log\, (\sigmagas/M_\odot \, \rm pc^{-2}) \approx 0$. We overplot the K-S relation derived by \cite{delosReyes2019} for local star-forming spiral galaxies (Figure~\ref{fig_KSlaw_compare}a) and spiral plus dwarf galaxies combined (Figure~\ref{fig_KSlaw_compare}b), which have slopes of 1.41 and 1.26, respectively. The relation between \sigmasfr\ and dust mass surface density (\sigmadust) for star-forming spiral galaxies from \cite{Kennicutt2021} has a similar slope (1.29; Figure~\ref{fig_KSlaw_compare}c). The shaded regions indicate the $1\sigma$ scatter of the corresponding relations, whose zero points have been adjusted to be consistent with the initial mass function of \cite{Chabrier2003} adopted in this work. All three relations intersect well the upper peak of star-forming galaxies in our sample. Although detailed comparisons are beyond the scope of the current paper, the overall qualitative consistency of our sample distribution with the K-S relations from previous works provides independent confirmation of the robustness of the parameters derived by \cite{Li2023}, who employed different methods to measure SFR, gas mass, and galaxy size. 

We note that the published K-S relations typically encompass minimum surface densities of only $\log\, (\sigmasfr/M_\odot \, \rm yr^{-1}\,kpc^{-2}) \approx -4$, $\log\, (\sigmagas/M_\odot \, \rm pc^{-2}) \approx 0$, and $\log\, (\sigmadust/M_\odot \, \rm pc^{-2}) \approx -2$. Our work extends the traditional K-S law to lower gas and dust surface densities by more than 1 order of magnitude and lower SFR surface densities by nearly 2 orders of magnitude. To our knowledge, previous observations never systematically investigated this regime of gas-poor, passive galaxies. Our measurements, obtained from SFRs and dust masses derived from fitting the FUV-to-FIR SEDs based on a new set of panchromatic photometry of galaxies in the Stripe~82 region, carry significant uncertainty for individual sources but no overall systematic bias \citep{Li2023}. Might the observed correlation between \sigmasfr\ and \sigmagas\ (\sigmadust) arise artificially from the fact that both SFR and \Md\ derive from SED fitting? We do not think so. To investigate this possibility, we replaced the SFRs from fitting the full (FUV--FIR) SED with less accurate SFRs based only on modeling the photometry of the five ($ugriz$) SDSS bands. The basic results shown in Figure~\ref{fig_KSlaw_compare} survive. 

We fit the entire sample with the orthogonal distance regression method in the {\tt Python} package \texttt{scipy.odr}.  This method is preferred over the simple $\chi^2$ method because the latter is strongly biased by the small fraction of targets on the lower right wing of the sample distribution.  The best-fit relation ($\alpha = 1.50$, $\beta = -4.29$) is slightly steeper than that from \cite{delosReyes2019}, such that the extrapolation of the latter deviates from our relation in the regime of passive galaxies at $\log\,(\sigmagas/M_{\odot}\,\rm pc^{-2})\lesssim 0$. We emphasize that our fitted relation mainly serves to facilitate qualitative comparison with the canonical K-S relation.  We opt not to fit the relation separately for the star-forming and passive galaxies because of their limited dynamic range and the considerable uncertainty of the gas masses. For completeness, we note that a similar deviation in the regime of passive galaxies appears in the $\sigmasfr - \sigmadust$ relation (Figure~\ref{fig_KSlaw_compare}c; \citealt{Kennicutt2021}).

\begin{figure*}
\centering
\includegraphics[width=0.86\textwidth]{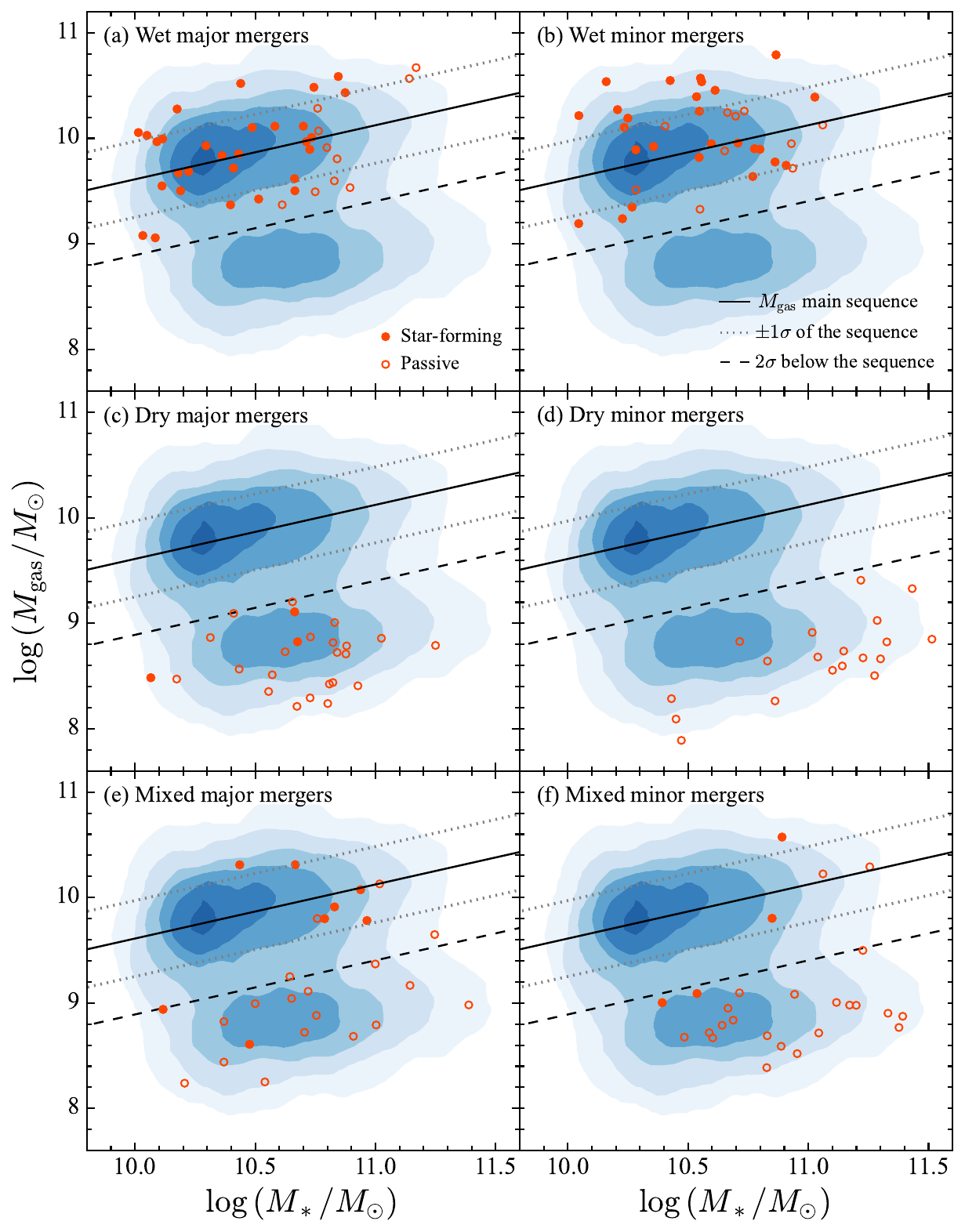}
\caption{Distribution of gas mass and stellar mass for the entire sample (background contours) and for the different categories of mergers: (a) wet major mergers, (b) wet minor mergers, (c) dry major mergers, (d) dry minor mergers, (e) mixed major mergers, and (f) mixed minor mergers. The primary galaxy of each pair is highlighted as a filled red point if it is star-forming and as an open red point if it is passive, which is defined by $\rm log\,(sSFR/yr^{-1})<-11$. The black solid line indicates the best fit for the gas mass main sequence, which has a $1\sigma$ scatter of 0.38~dex (gray dotted lines). The threshold for gas-rich objects is given by the black dashed line, which is set to $2\sigma$ below the gas mass main sequence.}
\label{fig_MgasMs}
\end{figure*}

\begin{figure*}
\centering
\includegraphics[width=0.86\textwidth]{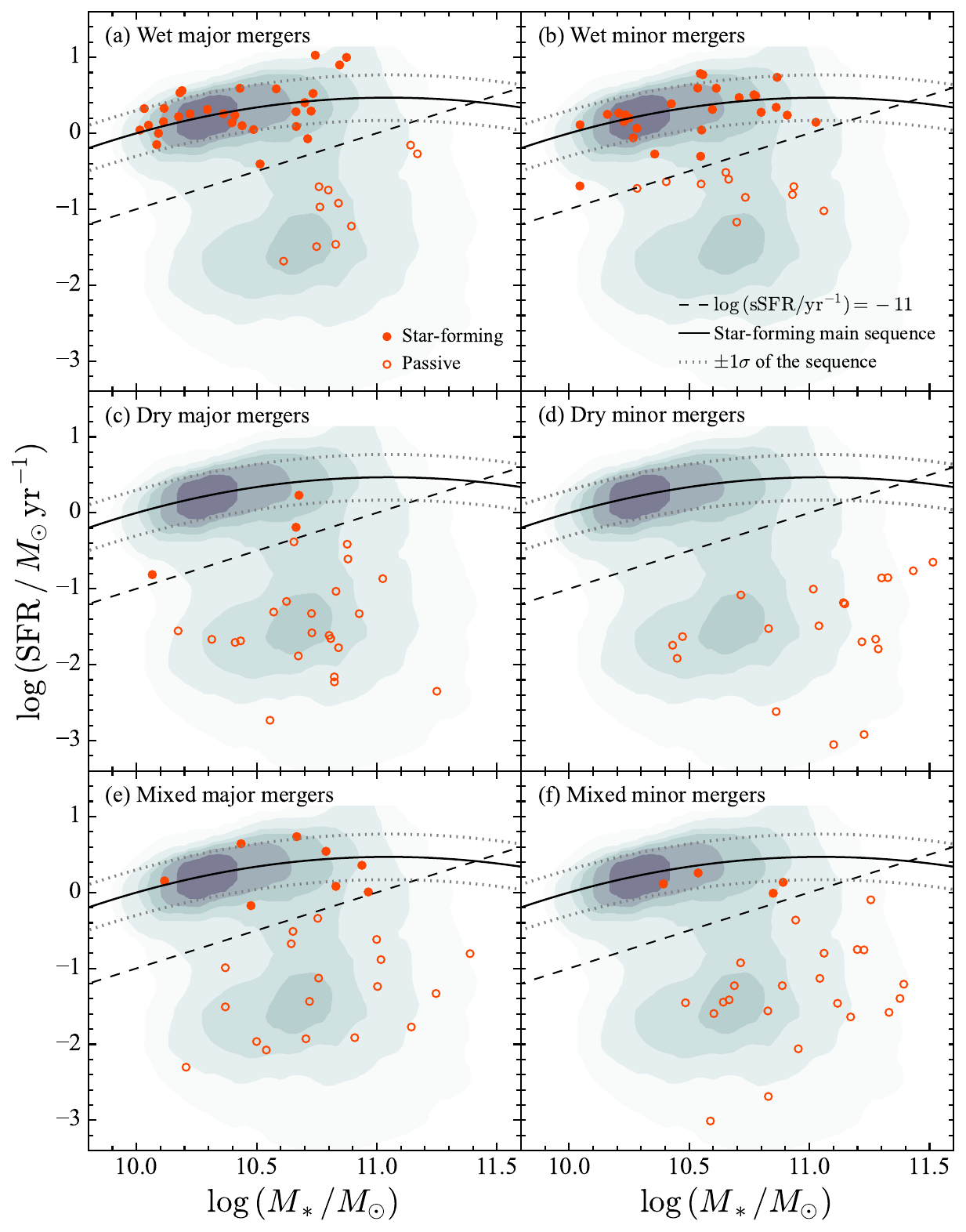}
\caption{Distribution of SFR and stellar mass for the entire sample (background contours) and for the different categories of mergers: (a) wet major mergers, (b) wet minor mergers, (c) dry major mergers, (d) dry minor mergers, (e) mixed major mergers, and (f) mixed minor mergers. The primary galaxy of each pair is highlighted as a filled red point if it is star-forming and as an open red point if it is passive, which is defined by $\rm log\,(sSFR/yr^{-1})<-11$. The black solid line indicates the best fit for the star-forming galaxy main sequence, which has a $1\sigma$ scatter of 0.3~dex (gray dotted lines). The black dashed line separates star-forming galaxies from passive galaxies.}
\label{fig_SFRMsDiagram}
\end{figure*}

\subsection{Gas Mass and SFR of Merger Galaxies}\label{sec_merger_prop}

Figure~\ref{fig:histAll} compares the distributions of SFR, sSFR, SFE, and $f_\mathrm{gas}$ for the merger and non-merger groups. The two-sample Kolmogorov-Smirnov test\footnote{\url{https://docs.scipy.org/doc/scipy/reference/generated/scipy.stats.kstest.html}} rejects the null hypothesis that SFR, sSFR, and $f_{\rm gas}$ are drawn from the same parent distribution, with a probability of $p = 0.005$, 0.003, and $<0.001$, respectively. Among the population of less actively star-forming galaxies, mergers have somewhat higher SFR and sSFR but lower $f_{\rm gas}$ than non-mergers. These then conspire to produce statistically indistinguishable SFEs for the two populations ($p = 0.679$).

To facilitate the following discussion, we define star-forming galaxies as objects with $\mathrm{sSFR}>10^{-11}\,\mathrm{yr^{-1}}$, and the rest as passive galaxies. We further classify the galaxies into gas-rich and gas-poor varieties, according to their distance from the gas mass main sequence, as given by Equation~3. Galaxies located more than $2\sigma$ below the gas mass main sequence are deemed gas-poor, while the rest are gas-rich. Consistent with common practice, we designate the merger of two gas-rich galaxies a ``wet merger,'' that of two gas-poor galaxies a ``dry merger,'' and the combination of a gas-rich and a gas-poor galaxy a ``mixed merger.'' We also separate mergers into major and minor categories based on whether their stellar mass ratio $f_{M_*} > 0.25$. In this study, we only investigate the more massive, primary component of the galaxy pair, such that it always meets the stellar mass cut of $M_* \ge 10^{10}\,M_\odot$, in order that the non-merger members in our parent sample can serve as control galaxies. If both components of the galaxy pair have comparable stellar masses within their $3\sigma$ measurement uncertainty of 0.15~dex (Section~\ref{sec_sample}), we include both in the merger sample.

\subsubsection{Mergers on the Gas Mass Main Sequence}\label{sec_gmms}

We examine the nature of merging galaxies on the gas mass main sequence of the entire sample, splitting the mergers by their gas content and stellar mass ratio. Most (74\%) of the wet major mergers are star-forming galaxies (Figure~\ref{fig_MgasMs}a). However, interestingly, the remaining 26\%, though massive and gas-rich, are passive. This phenomenon is not unique to the merger sample, for a nearly identical fraction (19\%) of the gas-rich non-mergers are also passive. Very few sources---only six, or $23\%$---actually lie $1\sigma$ above the gas mass main sequence, which indicates that wet major mergers before coalescence do not contain more cold gas than non-mergers. This is consistent with the previous finding that only late-stage mergers show enhanced cold interstellar medium \citep{Shangguan2019}. Wet minor mergers behave similarly but not identically (Figure~\ref{fig_MgasMs}b). A higher fraction of galaxies (33\%, almost double the fraction among gas-rich non-mergers) lie $1\sigma$ above the gas mass main sequence. A comparable fraction (28\%) of the wet minor merging galaxies are passive, moderately higher than the fraction of passive galaxies in non-merger galaxies. As expected, dry mergers are dominated almost exclusively by passive, gas-poor galaxies, with a tendency for systematically higher stellar masses in minor mergers (Figure~\ref{fig_MgasMs}d) compared to major mergers (Figure~\ref{fig_MgasMs}c). Mixed major mergers consist of  $\sim\,$60\% gas-poor galaxies (Figure~\ref{fig_MgasMs}e), similar to the fraction of gas-rich objects, while gas-poor systems account for the majority (85\%) of mixed minor mergers (Figure~\ref{fig_MgasMs}f). The implications of these results are discussed in Section~\ref{sec_discuss_lackSB}.

\vspace{0.5cm}
\subsubsection{Mergers on the SFR$-M_*$ Diagram}\label{sec_sfms}

We also compare the merger galaxies with the entire galaxy sample on the SFR$-$\Ms\ diagram (Figure~\ref{fig_SFRMsDiagram}). Irrespective of stellar mass ratio, wet mergers largely, but not exclusively, reside along the star formation main sequence. However, a significant fraction ($\sim 25\%-30\%$) occupies the so-called green valley, despite the ready supply of fuel available for star formation. Gas-poor systems are passive, as expected, while among mixed pairs a considerably larger fraction of objects form stars in major mergers than minor mergers. Most unexpected, perhaps, is the rarity of ``starbursts,'' defined here as galaxies that scatter at least $1\sigma$ (0.3~dex) above the star-forming main sequence \citep{Elbaz2007,Ellison2020,Wang2022}. Among wet mergers, be they major or minor, only 7 sources or 9\% of the total qualify as starbursts. This low fraction is almost the same as that for the control sample of non-mergers (8\%).

\begin{figure*}
\centering
\includegraphics[width=0.86\textwidth]{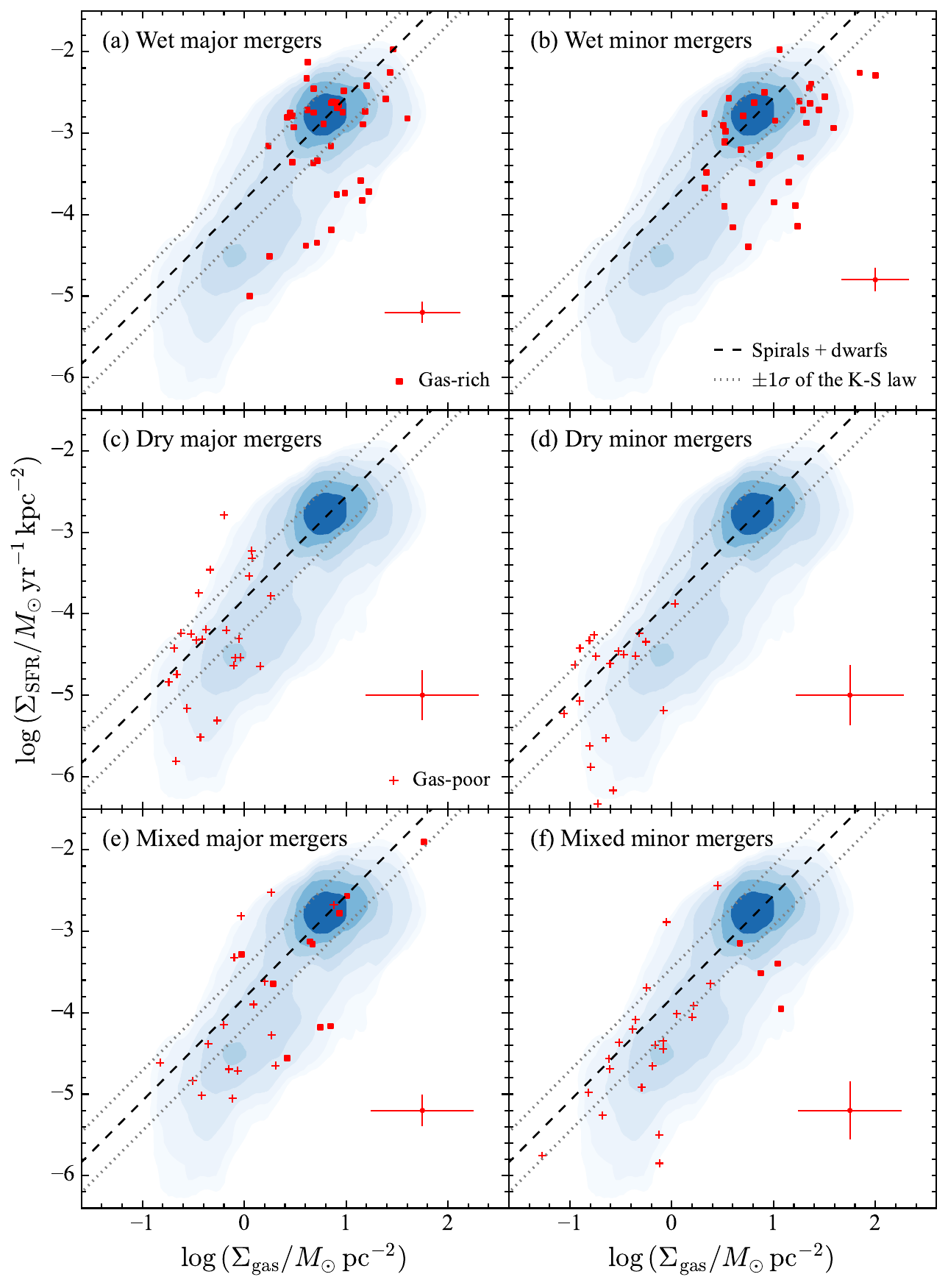}
\caption{Distribution of merger pairs on the \sigmasfr--\sigmagas\ plane, with the full sample shown as background contours, for the different categories of mergers: (a) wet major mergers, (b) wet minor mergers, (c) dry major mergers, (d) dry minor mergers, (e) mixed major mergers, and (f) mixed minor mergers. The typical uncertainties for each subsample of mergers are shown in the bottom-right corner of each panel. The black dashed line gives the K-S law for spiral and dwarf galaxies from \cite{delosReyes2019}, which has a $1\sigma$ scatter of 0.37~dex (gray dotted lines).}
\label{fig_KSlaw_mergers}
\end{figure*}

\begin{figure}
\centering
\includegraphics[width=0.492\textwidth]{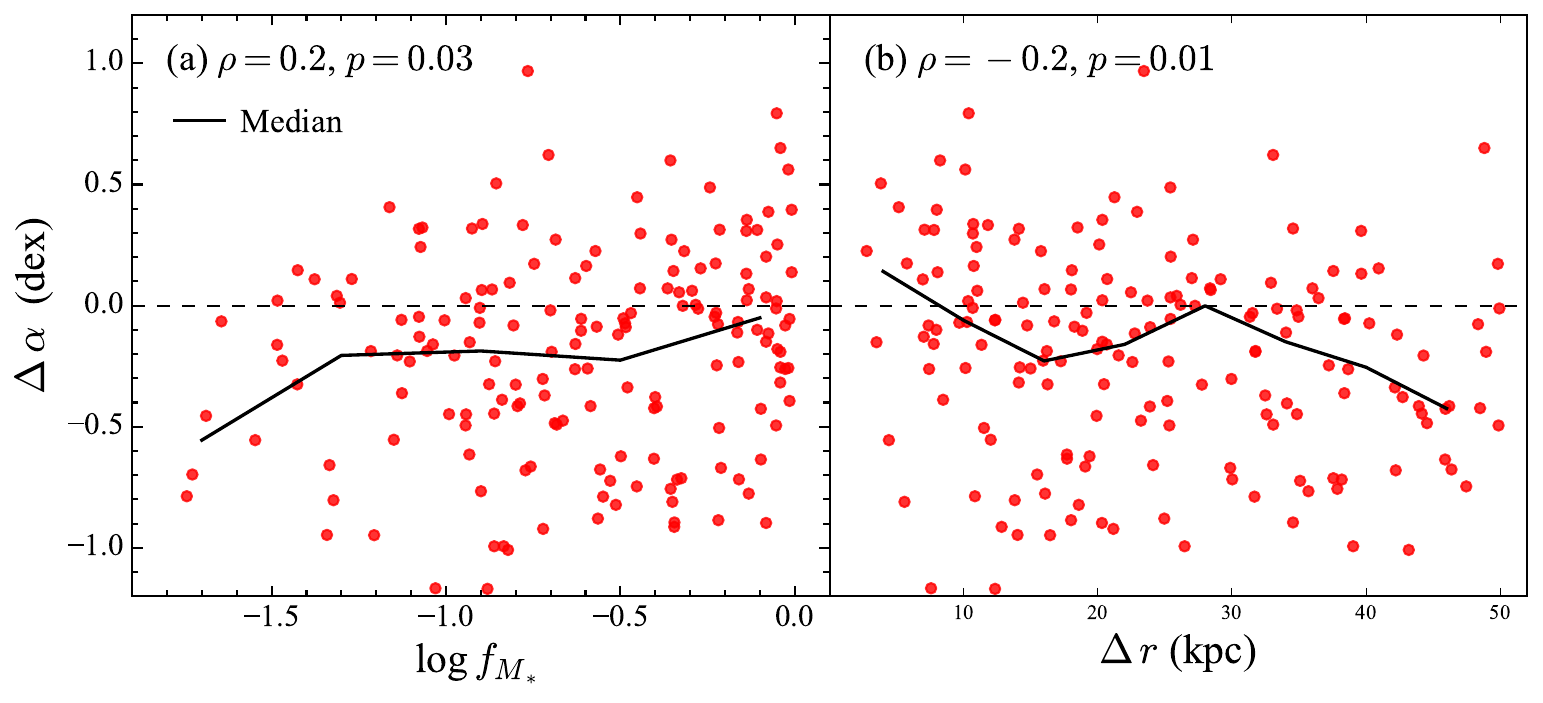}
\caption{Distance to the K-S law ($\Delta \alpha$) of \cite{delosReyes2019} for each galaxy in the sample of mergers, including both major and minor pairs, as a function of (a) the stellar mass ratio ($f_{M_*}$) and (b) projected separation ($\Delta r$). The Spearman's rank correlation coefficient $\rho$ and $p$-value are shown in the upper-left corner of each panel. A positive correlation has $\rho>0$, and a negative correlation has $\rho<0$. The values of $p<0.05$ and $|\rho| \approx 0.2$ imply that there is a weak correlation between the two parameters. The black line indicates the average trend in each panel.}
\label{fig_H_wetdry}
\end{figure}

\subsubsection{Mergers on the Kennicutt-Schmidt Relation}\label{sec_KS_merger}

Finally, we investigate the behavior of mergers on the \sigmasfr$-$\sigmagas\ plane. To first order, the overarching impression from Figure~\ref{fig_KSlaw_mergers} is that mergers in the local Universe leave, at best, a rather weak imprint on the efficiency of star formation of galaxies. Consistent with a theme echoed throughout this study, we find little evidence that mergers significantly boost the SFE, certainly not to the extent expected of starburst  galaxies \citep{Daddi2010b, Genzel2010, Saintonge2011, Kennicutt2021}. Similar conclusions are also reached by \citet{DiazGarcia2020} and \citet{Violino2018}, who, using a different approach from ours, found that the gas depletion time of interacting galaxies is not statistically different from that of non-interacting galaxies. Our conclusion applies to mergers with varying degrees of gas-richness and stellar mass ratio. Only two galaxies in the wet major merger subsample genuinely lie above the 0.37 dex ($1\sigma$) scatter of the K-S relation of \cite{delosReyes2019} for spiral and dwarf galaxies. On the contrary, some galaxies in wet major mergers even lie well {\it below}\ the K-S relation. This effect is most pronounced for the sample of wet minor mergers, whose entire sample exhibits a pronounced, systematic offset toward lower SFE. The existence of gas-rich yet passive galaxies is puzzling. Note that these results remain qualitatively unchanged if we substitute other published K-S relations as reference instead.

What factors drive the residual offset from the K-S law?  Figure~\ref{fig_H_wetdry} shows that among mergers, $\Delta\alpha$ with respect to the K-S law of \cite{delosReyes2019} systematically increases with increasing stellar mass ratio ($f_{M_*}$; panel a) and closer pair separation ($\Delta r$; panel b). We emphasize that, despite the formal significance of the statistical test, the correlations are only moderate and the scatter significant. In galaxy pairs of comparable stellar mass (high log$\,f_{M_*}$) the SFR can be enhanced during the late stage (small $\Delta\,r$) of the merger, presumably on account of the strong tidal and turbulent compression of the gas. The large scatter of $\Delta \alpha$ is due to the large uncertainty in the gas mass, propagated from the uncertainty of the dust mass and $\delta_{\rm GDR}$. Those weak trends demonstrate that the increase of SFE is not always effectively triggered by mergers. 

\vspace{1.5cm}
\section{Discussion}\label{sec_discussion}

\subsection{Extension of the Kennicutt-Schmidt Relation to Quiescent Galaxies}
\label{sec_discuss_KS}

Our improved photometric measurements and SED analysis have greatly expanded the dynamical range over which SFR and dust (gas) mass can be measured in low-redshift galaxies. In addition to the detailed quantification of uncertainties and other technical improvements, our treatment of the upper limits for the Herschel FIR data turns out to be critical to derive unbiased SFRs and dust masses for the passive galaxies \citep{Li2023}. Although the measurements are quite uncertain for individual sources, their sample distribution is robust and still highly informative. For example, our sample reaches ${\rm SFR} \approx 10^{-2}\,M_\odot\,\mathrm{yr^{-1}}$, an order of magnitude lower than the values routinely obtained from SED fits that only include FIR detections (e.g., Rowlands et al. 2012, 2014). Eales et al. (2017) report sSFR~$\approx 10^{-12.5}\,\mathrm{yr^{-1}}$ for the early-type galaxies from HRS (distance 15--25~Mpc), which, with $M_*\approx 10^{10.5}\,M_\odot$, closely resemble the passive galaxies in our sample. It is not surprising, then, to discover some low-level star formation in our passive galaxies. Many observations show that massive early-type galaxies can retain a large amount of atomic or molecular gas (e.g., \citealt{Combes2007,Babyk2019}), even though they are on or transiting to the red sequence \citep{Young2014}. 

This study mainly covers star-forming and passive galaxies, with very few starburst systems, owing to the unbiased nature of our sample selection. Not surprisingly, then, the $\sigmasfr-\sigmagas$ relation of the star-forming subsample (Figure~\ref{fig_KSlaw_myfit}) agrees well with published results (e.g., \citealt{Kennicutt1998,Kennicutt2021}): the slope of $\beta = 1.50$ lies comfortably within the $1-2\sigma$ range of previous studies \citep{Leroy2008, Daddi2010a, Martig2013, Liu2015}. When extending to passive galaxies, the best-fitting slope deviates from the extrapolation of the canonical relation based on spiral and dwarf galaxies. The power-law index for the K-S relation may differ between star-forming and starburst galaxies, and it depends on whether the gas measurement includes total gas or just molecular gas. For example, while \cite{Kennicutt2021} found a break in the slope of the K-S relation that involves total gas density; when molecular gas is used, the relation is roughly linear for all but the very lowest surface density systems, which fall below a threshold density \citep{delosReyes2019}. The presence of a threshold gas density is also suggested by \citet{Bigiel2008}, who found that star formation becomes inefficient when the local resolved gas surface density $\sigmagas\lesssim 10\,M_{\odot}\,\rm pc^{-2}$. At low gas surface density, the interstellar medium resides mainly in atomic instead of molecular phase, which leads to a steeper K-S relation \citep{Leroy2008, Bigiel2014, delosReyes2019}.

Within our sample, a possible break in the \sigmasfr--\sigmagas\ relation occurs in the transition region between star-forming and passive galaxies (Figure~\ref{fig_KSlaw_myfit}). The star-forming galaxies are mostly located above $\sigmagas \approx {\rm few} \,M_{\odot}\,\rm pc^{-2}$, which is intriguingly close to the threshold of $\sigmagas \approx 10\,M_{\odot}\,\rm pc^{-2}$ discussed by \cite{Bigiel2008}. By contrast, the passive galaxies populate the region of lower density, at $\sigmagas \approx 0.1-1\, M_{\odot}\,\rm pc^{-2}$. The numerical simulations of \cite{Becerra2014} suggest that the $\sigmasfr-\sigmagas$ relation may extend as far as $\sigmagas \approx 10^{-3}\,M_{\odot}\,\rm pc^{-2}$, with a steepening of the slope at $\sigmagas\lesssim 1 \,M_{\odot}\,\rm pc^{-2}$, in good qualitative agreement with our observations.

\begin{figure}
\centering
\includegraphics[width=0.43\textwidth]{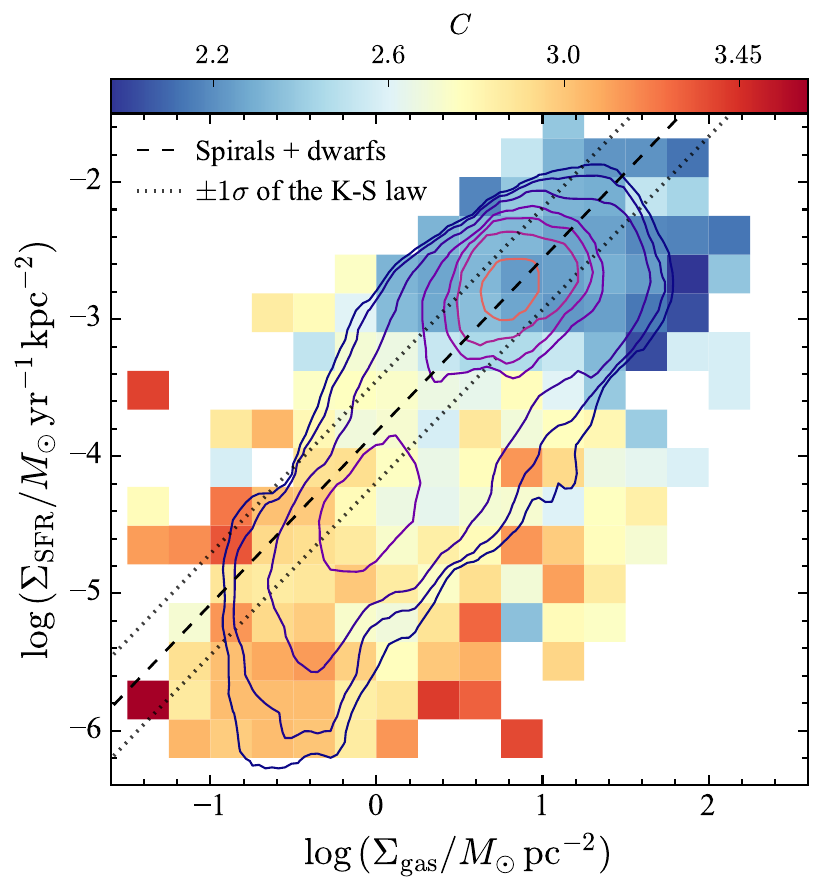}
\caption{K-S relation for our sample (contours), color-coded by the average concentration index $C \equiv R_{90}/R_{50}$ in two-dimensional bins of \sigmasfr\ and \sigmagas, with $R_{90}$ and $R_{50}$ the \cite{Petrosian1976} radii enclosing 90\% and 50\% of the $r$-band light, respectively. The black dashed line gives the relation for spiral and dwarf galaxies from \cite{delosReyes2019}, which has a $1\sigma$ scatter of 0.37~dex (gray dotted lines).}
\label{fig_KS_C}
\end{figure}

Why do passive galaxies have a steeper \sigmasfr--\sigmagas\ relation? On the one hand, the superlinear relation between SFR and total gas surface density can be shaped by the combination of a nearly linear molecular gas relation and the almost vertical trend between SFR and atomic gas surface density \citep{Bigiel2008, Leroy2013}. Since the molecular gas mass fraction in the interstellar medium drops significantly when $\sigmagas\lesssim 10\,M_{\odot}\,\rm pc^{-2}$ \citep{Wong2002, Saintonge2022}, massive early-type galaxies may be dominated by \HI\ gas \citep{Young2014, Saintonge2016}, which may not be conducive to forming stars \citep{McKee2007, Krumholz2014}. \cite{Verley2010} observed a steep \sigmasfr--\sigmagas\ relation for the \HI-dominated regions of M\,33. On the other hand, the stability of a gas disk in a star-forming galaxy may be quite different from that of a passive galaxy \citep{Martig2009}, in view of the much higher concentration (bulge-dominance) of passive galaxies compared to star-forming galaxies (Figure~\ref{fig_KS_C}). High concentration produces a steeper gravitational potential well and hence stronger shear in the velocity field, which, in turn, increases Toomre's (1964) $Q$ parameter and prevents the cold gas from collapsing to form stars \citep{Krumholz2005Natur, Somerville2015}. Gas-rich, early-type galaxies can have SFEs lower by a factor of more than 2 compared to late-type galaxies at fixed gas surface density \citep{Martig2013, Davis2014}. Interestingly, the concentration parameter almost exactly demarcates the two populations of star-forming and passive galaxies in Figure~\ref{fig_KS_C}.

As a caveat, we emphasize that the SFRs and gas masses of individual passive galaxies carry considerable uncertainties. Although \cite{Li2023} show that the sample distribution of the measured physical quantities is unbiased, the large uncertainties of the SFR and gas mass estimates can affect the sample distribution and bias the slope of the $\sigmasfr - \sigmagas$ relation for the star-forming and passive populations. This effect is unlikely to affect the scaling relation of the entire sample, given the large dynamic range of the overall distribution. However, it may bias the slopes if one fits the star-forming and passive subsamples separately. Moreover, we estimated the gas mass using the dust mass and the mass--metallicity relation derived from star-forming galaxies \citep{KewleyEllison2008, Magdis2012}. While we use different stellar mass measurements than the estimates used by \cite{KewleyEllison2008}, it is essential to adopt the $\delta_{\rm GDR}-\rm Z$ and $M_*-\rm Z$ relations self-consistently in terms of the metallicity calibration \citep{Berta2016}. Different calibrations can lead to significant systematic discrepancies for the $M_*-\rm Z$ relation in terms of both its shape and scale \citep{KewleyEllison2008}. In this work, estimations from Equations~\ref{equ_GDR} and \ref{equ_metal} are both based on the calibration of \cite{PettiniPagel2004} at $M_*\approx 10^{11}\,M_{\odot}$. The mass--metallicity relation may differ in passive galaxies. \cite{Wu2020ApJ} argues, for instance, that the gas-phase metallicity of early-type galaxies is $\sim 0.1$~dex lower than in later-type, star-forming galaxies, possibly as a result of dilution by pristine gas from the intergalactic medium. It is beyond the scope of the current work to investigate this effect. We simply note that a lower metallicity in passive galaxies would correspond to a higher gas-to-dust ratio and, thus, a higher gas mass, which may lead to an even steeper $\sigmasfr - \sigmagas$ relation.

\subsection{The Low Frequency of Starbursts in Mergers}\label{sec_discuss_lackSB}

Only nine (5\%) of the 174 primary galaxies in the merger sample qualify as starburst galaxies, by the commonly adopted criterion that they should lie at least $1\sigma$ ($>0.3$~dex) above the star-forming main sequence \citep{Elbaz2007, Ellison2020, Wang2022}. This starburst fraction is similar to that in the parent sample, which is dominated vastly by non-mergers. Visual inspection reveals that these nine starburst galaxies all have late-type structures, most showing obvious signs of morphological disturbance. Moreover, all have high gas fractions and have a gas-rich companion. Six of the nine can be considered major mergers, characterized by a median stellar mass ratio of $\sim 0.7$. Our results, therefore, are consistent with previous observations and simulations, which conclude that  starburst galaxies, if triggered by galaxy mergers, are primarily associated with gas-rich major mergers, instead of other types of merger pairs \citep{Kartaltepe2012, Hung2013, Patton2013, Pan2018}. Strong tidal torques imparted by the merger of galaxies of comparable mass effectively promote gas inflow and subsequent vigorous star formation \citep{Mihos1996, Cox2006, Renaud2014, Sparre2022}. 

At first glance, the absence of an increased incidence of starbursts in mergers seems to conflict with previous works. For instance, luminous and ultraluminous infrared galaxies (LIRGs and ULIRGs) mostly consist of galaxies in different merger stages \citep{Sanders1996, Kaviraj2009, Espada2018, He2022BTr}. So-called submillimeter galaxies, the higher redshift counterparts of the more nearby ULIRGs, are strong starburst systems with SFR up to $1000\,M_{\odot}\,\rm yr^{-1}$ \citep{Tacconi2006} that are often affiliated with major mergers of gas-rich disks (\citealt{Engel2010, Michalowski2010}; but see \citealt{Cheng2022}).  However, we find that the reverse does not hold. Major mergers, even when gas-rich, are not necessarily starbursts (Figure~\ref{fig_MgasMs}a).  Numerical simulations demonstrate that while the tidal torque of a major merger drives cold gas to the center, gas simultaneously also gets ejected to the outskirts of the galaxy in the form of tidal tails \citep{DiMatteo2007}. The outer regions of interacting systems may store a significant fraction of the gas ejected from the inner regions of the galaxies \citep{Perez2011}. The gas at large radii only partially flows back to the center at late times. The gas inflow depends on the details of the galaxy encounter, such as the orientation, pericentric distance, and the size of the gas disk \citep{Blumenthal2018}. In addition to the well-known gas inflow, other mechanisms, such as the interaction-triggered gas migration \citep{Pan2019}, increased gas turbulence \citep{Pan2018, Violino2018}, or the formation of internal structure in the galaxy \citep{Espada2018} can regulate the level of star formation in the merging components. \citet{Pan2019} investigate the radial distribution of SFR derived from \Ha\ emission for merging galaxies and a control sample. They find the radial profiles of sSFR, which can be enhanced or suppressed, have considerable variance across the merger stages, while the global sSFR before the coalescence phase is not significantly enhanced.  The probability of witnessing a merger undergoing a starburst episode depends on when we happen to catch it during its evolution,  as well as sample selection bias \citep{Murphy1996}. Furthermore, (U)LIRGs are very rare in the local Universe \citep{Soifer1991, Knapen2009}. Most (U)LIRGs are discovered during the merging of two disk galaxies \citep{Colina2001, Carpineti2015}, which constitute only a small fraction of the entire merger sample. Studying a large sample of the nearest galaxies, \citet{KnapenCisternas2015} find that starbursts only show a moderate preference for interacting galaxies, and that the results strongly depend on sample selection.
    
Wet minor mergers present an even more extreme situation with their apparently suppressed SFE (Figure~\ref{fig_KSlaw_mergers}b). At fixed \sigmagas, wet minor mergers have systematically lower \sigmasfr\ compared to non-mergers. A few outliers with significantly suppressed SFE can also be found among the wet major mergers. Figure~\ref{fig_KS_C} offers a possible explanation for this otherwise somewhat perplexing behavior. Some of the most extreme outliers below the K-S relation have relatively high concentration, reminiscent of dust-rich early-type galaxies, which, having experienced a recent minor merger now forms stars inefficiently because the gas is still streaming toward the center or the interstellar medium is heated by shocks and turbulence \citep{Davis2015}.  The huge scatter of mergers on the K-S relation confirms the suggestion 
of \citet[their Figure 4]{DiazGarcia2020} that gas mass enhancement barely correlate with SFR enhancement of SFR. Other recent observations arrive at similar conclusions regarding the mild effect of mergers on the global SFR of galaxies \citep[e.g.,][]{Puech2014, Silva2018}. Mergers and non-mergers occupy almost indistinguishable positions on the star-forming main sequence at $z \approx 0-4$ \citep{Xu2010, Wong2011, Lackner2014, Knapen2015, Silva2018, Pan2019, Pearson2019, DiazGarcia2020}. Only $\sim 12\%$ of  massive, closely separated ($\Delta r = 3-15$~kpc) major mergers at $0.3<z<2.5$ are starbursts \citep{Silva2018}. This fraction is higher than ours. Besides the difference in redshift, our work in Stripe~82 detects more passive galaxy components in dry and mixed mergers, which represent 27\% and 30\% of all mergers in our sample, respectively. If we take this into consideration, the fraction of starburst galaxies in our merger sample becomes even lower. 

This work mainly focuses on the primary galaxy component of each merger pair. Star formation in the less massive member of the pair may be impacted more dramatically by tidal compression or gas stripping \citep{DiMatteo2007, Silva2018, Du2019}. The detailed study of the secondary companion is beyond the scope of this work.

\section{Summary}\label{sec_summary}

The recently completed catalog of \cite{Li2023} provides uniformly measured panchromatic photometry and physical parameters of 2685 massive ($M_* \ge 10^{10}\,M_\odot$), low-redshift ($z = 0.01-0.11$) galaxies selected from the SDSS Stripe~82 region. In addition to stellar masses and SFRs, the new analysis supplies total (atomic and molecular hydrogen) gas masses inferred from total dust masses derived from careful analysis of the full broadband (FUV--FIR) SED. The accurate measurements of \cite{Li2023}, in combination with their SED analysis that fully incorporates proper uncertainties and upper limits, yield an exceptionally wide dynamic range in parameter space. Our catalog reaches ${\rm SFRs} \approx 10^{-2}\,M_\odot\,\mathrm{yr^{-1}}$ and $M_{\rm gas} \approx 10^{8.5}\,M_\odot$, which enables us to probe the unbiased star formation activity and gas content of the full population of massive galaxies, including previously poorly studied gas-poor, largely quiescent members.

We take advantage of this resource to investigate the effect of galaxy interactions and mergers on star formation. Selecting merging galaxies in a relatively advanced stage of interaction (projected separation $\Delta\,r<50$~kpc), we study the influence of late-stage mergers on the gas and star formation properties, comparing them to the dominant, control sample of non-merger galaxies within the framework of the well-established galaxy scaling relations between SFR and $M_*$ on the one hand, and between \sigmasfr\ and \sigmagas\ on the other hand. We consider the effect of stellar mass ratio by categorizing the mergers as major or minor, and the impact of gas fraction by defining them as wet, mixed, or dry from their position relative to the gas mass main sequence on the \Mg$-$\Ms\ diagram.

Our main results can be summarized as follows:

\begin{enumerate}
\item
The distribution of galaxies in the \Mg$-$\Ms\ plane is bimodal. Apart from the gas-rich, mostly star-forming galaxies that trace the gas mass main sequence, we identify a clearly distinct, secondary population of gas-poor, largely quiescent galaxies.

\item
The bimodality of galaxies in the \Mg$-$\Ms\ plane maps to a similar bimodality in the \sigmasfr$-$\sigmagas\ plane. The entire population can be described by $\sigmasfr \propto \sigmagas^{1.50}$, not dissimilar from previous studies of the Kennicutt-Schmidt relation, although our sample extends to considerably more passive, gas-poor galaxies that exhibit a steeper power-law relation compared to the conventionally studied star-forming, gas-rich systems.

\item
Low-redshift galaxy mergers show very minimal differences in their star formation properties (SFR, sSFR, SFE) compared to non-mergers, although the SFE of the galaxies in mergers varies mildly with stellar mass ratio and pair separation. Contrary to popular wisdom, we find no excess of starbursts in gas-rich, major mergers. Gas-rich, minor mergers, if anything, form stars even less efficiently than non-merging galaxies. 

\item
Our study highlights the subtle, even counterintuitive, impact of galaxy interactions and mergers on star formation activity in galaxies. Whether mergers convert gas into stars effectively or not depends on many factors, including the phase of the orbital evolution of the merging pair and the location of the gas as it responds to the tidal field of the gravitational interaction. 
\end{enumerate}

\begin{acknowledgments}
We thank the anonymous referee for helpful comments. This work was supported by the National Science Foundation of China (11721303, 11991052, 12011540375, 12233001), the National Key R\&D Program of China (2022YFF0503401), and the China Manned Space Project (CMS-CSST-2021-A04, CMS-CSST-2021-A06). We thank the following individuals for their discussions and contributions to various aspects of this work: Ruancun Li, Chao Ma, Yingjie Peng, Longfei Zhai, Dingyi Zhao, and Ming-Yang Zhuang.
\end{acknowledgments}

\end{document}